\documentclass[floatfix,aps,amsmath,nofootinbib,twocolumn,10pt]{revtex4}

\usepackage{listings}
\usepackage{graphicx}
\usepackage{bm}
\usepackage{rotating}
\usepackage{array}
\usepackage{amsmath}
\usepackage{amssymb} 
\usepackage{mathrsfs} 
\usepackage{cancel}

\def\({\left(}
\def\){\right)}
\def\[{\left[}
\def\]{\right]}

\def\e{\begin{equation}}
\def\q{\end{equation}}
\def\m{\begin{eqnarray}}
\def\n{\end{eqnarray}}

\begin{document}

\title{Gravitational Waves from Preheating with Modified Gravitational-wave Propagation}

\author{Ke Wang}
\email{wangkey@lzu.edu.cn}
\affiliation{Institute of Theoretical Physics \& Research Center of Gravitation,\\ Lanzhou University, Lanzhou 730000, China}
\author{Qing-Guo Huang}
\email{huangqg@itp.ac.cn}
\affiliation{CAS Key Laboratory of Theoretical Physics, 
Institute of Theoretical Physics, Chinese Academy of Sciences,
Beijing 100190, China}
\affiliation{School of Physical Sciences, 
University of Chinese Academy of Sciences, 
No. 19A Yuquan Road, Beijing 100049, China}
\affiliation{School of Fundamental Physics and Mathematical Sciences
Hangzhou Institute for Advanced Study, UCAS, Hangzhou 310024, China}
\affiliation{Center for Gravitation and Cosmology, College of Physical Science and Technology, Yangzhou University, Yangzhou 225009, China}

\date{\today}

\begin{abstract}

We investigate how the production of gravitational waves (GWs) is affected by the GW velocity $(c_T)$ during preheating after inflation. For instance, we simulate the production of GWs after  $\lambda\phi^4$ chaotic inflation, and find that GW spectrum is enhanced for $c_T<1$, but distorted (suppressed at low frequency, but enhanced at high frequency) for $c_T>1$. 


\end{abstract}

\pacs{???}

\maketitle


\section{Introduction}
\label{sec:intro}
In principle, during inflation, the quantum fluctuations of spacetime can generate a stochastic gravitational wave background (GWB) with a near scale-invariant spectrum. Since the amplitude of this spectrum is proportional to the energy scale of inflation, it can be used to probe the inflation scale and measured by the cosmic microwave background (CMB) B-mode. Combing current Plank data with BICEP2/Keck 2015 date gives an upper limit of tensor/scalar ratio $r<0.044$ at $95\%$ confidence~\cite{Tristram:2020wbi}. After the end of inflation, inflaton field will oscillate homogeneously at the bottom of the inflationary potential. If parametric resonance (or preheating)~\cite{Traschen:1990sw,Kofman:1994rk} occurs, the amplitudes of inflaton field grow exponentially and the growth rate is dominated by the Mathieu characteristic exponent. Usually, the amplitudes of inflaton field grow faster for smaller $k$. As a result, there is a large inhomogeneity in the distribution of fields and their energy densities. Therefore, a GWB would be produced at the corresponding wave numbers in a more ``classical" manner at the same time.

The production of GWs during preheating after inflation has been studied for many different inflation models~\cite{Easther:2006gt,Easther:2006vd,GarciaBellido:2007af,Dufaux:2007pt,Easther:2007vj,Huang:2011gf,Giblin:2011yh,Zhou:2013tsa,Ashoorioon:2013oha,Liu:2017hua,Amin:2018xfe,Adshead:2018doq,Liu:2018rrt,Sang:2019ndv,Jin:2020tmm,Hiramatsu:2020obh,Sang:2020kpd}. It's worth pointing out that there was an assumption of the validity of general relativity (GR) when these studies were done. Undoubtedly, if there are some modifications on GR, the resonance structure is expected to be different and then put some fingerprints in the GW power spectrum. 
Here, for simplicity, we will focus on a specific modification which doesn't affect the evolution of inflaton field and background, but change the propagation of tensor perturbations, i.e. GW propagation speed $c_T$ can be different from speed of light.

Our paper is organized as follows. In section~\ref{sec:model}, we set up the model which is considered in this paper. In section~\ref{sec:results}, we use PSpectRe to simulate the production of GWs from preheating after $\lambda\phi^4$ inflation. Finally, a brief summary and discussion are included in section \ref{sec:sum}.

In this paper, we adopt the following conventions: Greek indices run in $\{0, 1, 2, 3\}$, Latin indices run in $\{1, 2, 3\}$ and repeated indices implies summation and we are in natural units system with $\hbar=c=1$.

\section{The Model}
\label{sec:model}
A homogeneous and isotropic universe can be described by a flat Friedmann-Lemaitre-Robertson-Walker (FLRW) metric 
\begin{equation}
ds^2=-dt^2+a^2(t)(dx^2+dy^2+dz^2). 
\end{equation}
During the general hybrid inflation epoch, the evolution of the scale factor $a(t)$ is dominated by the Friedmann equations 
\begin{eqnarray}
\frac{\ddot{a}}{a}&=&-\frac{4\pi G_N}{3}\left( \rho(\phi,\chi)+3p(\phi,\chi)\right),\\
H^2&=&\left(\frac{\dot{a}}{a}\right)^2=\frac{8\pi G_N}{3}\rho(\phi,\chi),
\end{eqnarray}
where dots denote derivative with respect to physical time $t$, $\phi$ is the inflaton field and $\chi$ is a generic matter field. Therefore, the energy density $\rho(\phi,\chi)$ and press $p(\phi,\chi)$ are given by these fields' kinetic energy $T(\phi,\chi)$, gradient energy $G(\phi,\chi)$ and (generic quartic) potential energy $V(\phi,\chi)$
\begin{eqnarray}
\rho(\phi,\chi)&=&T(\phi,\chi)+G(\phi,\chi)+V(\phi,\chi),\\
p(\phi,\chi)&=&T(\phi,\chi)-\frac{1}{3} G(\phi,\chi)-V(\phi,\chi),\\
T(\phi,\chi)&=&\frac{1}{2}(\dot{\phi}^2+\dot{\chi}^2),\\
G(\phi,\chi)&=&\frac{1}{2a^2}((\nabla\phi)^2+(\nabla\chi)^2),\\
\nonumber
V(\phi,\chi)&=&\frac{1}{2}m_{\phi}^2\phi^2+\frac{1}{4}\lambda_{\phi}\phi^4\\
&+&\frac{1}{2}m_{\chi}^2\chi^2+\frac{1}{4}\lambda_{\chi}\chi^4+\frac{1}{2}g^2\phi^2\chi^2.
\end{eqnarray}
For the same period, the evolution of these fields obey the Klein-Gordon equation
\begin{eqnarray}
&\ddot{\phi}+3\frac{\dot{a}}{a}\dot{\phi}-\frac{1}{a^2}\nabla^2\phi+\frac{\partial V(\phi,\chi)}{\partial\phi}=0,\\
&\ddot{\chi}+3\frac{\dot{a}}{a}\dot{\chi}-\frac{1}{a^2}\nabla^2\chi+\frac{\partial V(\phi,\chi)}{\partial\chi}=0;
\end{eqnarray}
and the perturbed Einstein's equations give the evolution of the tensor perturbations $u_{ij}$
\begin{eqnarray}
\nonumber
&&\ddot{u}_{ij}+3\frac{\dot{a}}{a}\dot{u}_{ij}-\frac{1}{a^2}\nabla^2u_{ij}\\
\nonumber&=&\frac{16\pi G_N}{a^2}\bigg(\partial_i\phi\partial_j\phi+\partial_i\chi\partial_j\chi\\
&-&\left.\frac{1}{3}\delta_{ij}(\partial_k\phi\partial^k\phi+\partial_k\chi\partial^k\chi)\right).
\end{eqnarray}
From now on we switch to a specific modified gravity theory in which only the equation of motion (EOM) of tensor perturbations is changed. More precisely, we turn to a formulation of tensor perturbations' EOM with the tensor perturbations (or GWs) propagation speed $c_T\neq1$ in an effective field theory~\cite{Saltas:2014dha}, 
\begin{eqnarray}
\label{eq:meom}
\nonumber
&&\ddot{u}_{ij}+3\frac{\dot{a}}{a}\dot{u}_{ij}-\frac{c^2_T}{a^2}\nabla^2u_{ij}\\
\nonumber
&=&\frac{16\pi G_N}{a^2}\bigg(\partial_i\phi\partial_j\phi+\partial_i\chi\partial_j\chi\\
&-&\left.\frac{1}{3}\delta_{ij}(\partial_k\phi\partial^k\phi+\partial_k\chi\partial^k\chi)\right).
\end{eqnarray}
 
The transverse-traceless part of the tensor perturbations $h_{ij}$ can be extracted from $u_{ij}$ by the spatial projection operators~\cite{Misner:1973}  
\begin{eqnarray}
h_{ij}&=&S^{\rm TT}_{ij}u_{ij},\\
S^{\rm TT}_{ij}&=&P_{ik}S_{kl}P_{lj}-\frac{1}{2}P_{ij}(P_{lm}S_{lm}),\\
P_{ij}&=&\delta_{ij}-\frac{k_ik_j}{k^2}, 
\end{eqnarray}
and the energy-momentum tensor of GWs is given by 
\begin{equation}
T_{\mu\nu}=\frac{1}{32\pi G_N}\langle \partial_{\mu} h_{ij} \partial_{\nu}h^{ij}-\frac{1}{2}g_{\mu\nu}h_{ij}\square h^{ij}\rangle,
\end{equation}
where $\square$ is the d'Alembertain defined in the flat FLRW metric $g_{\mu\nu}$. Here $\langle \cdots \rangle$ denotes a spatial average which is taken over a region of large enough volume. Since we focus on a modified EOM of tensor perturbations, the propagation equation of $h_{ij}$ in vacuum is also  changed correspondingly, which means 
\begin{equation}
\square h_{ij}=\ddot{h}_{ij}+3\frac{\dot{a}}{a}\dot{h}_{ij}-\frac{1}{a^2}\nabla^2h_{ij}=\frac{c^2_T-1}{a^2}\nabla^2h_{ij}. 
\end{equation}
It is equal to zero for $c_T=1$, but becomes non-zero for $c_T\neq 1$. 
In order to compute the energy density of GWs, we need to compute the average over the lattice volume $L^3$. The energy density of GWs is just the $00$ component 
\begin{equation}
\rho_{\rm GW}=\frac{1}{32\pi G_N}\langle\dot{h}_{ij}\dot{h}^{ij}\rangle.
\end{equation}
Here, we found that the contribution from the term of $\frac{1}{2}h_{ij}\square h^{ij}$ can be ignored because of $\langle  h_{ij}\square h^{ij}\rangle \ll \langle\dot{h}_{ij}\dot{h}^{ij}\rangle $ for $\lambda\phi^4$ model in our simulations.
In momentum space, the energy density is
\begin{equation}
\rho_{\rm GW}=\sum_{i,j}\frac{1}{32\pi G_N}\frac{4\pi}{L^3}\int d\ln k~k^3|\dot{h}_{ij}(k)|^2, 
\end{equation}
and then the GW power spectrum per $\ln k$ normalized to the critical density during the evolution $\rho_c(a)$ is
\begin{equation}
\frac{d\Omega_{\rm GW}(a)}{d\ln k}=\frac{1}{\rho_c(a)}\frac{d\rho_{\rm GW}}{d\ln k}=\frac{\pi k^3}{3 H^2 L^3}\sum_{i,j}|\dot{h}_{ij}(k)|^2.
\end{equation}
Finally the GW power spectrum per $\ln k$ today is given by~\cite{Easther:2006gt,Easther:2007vj}
\begin{equation}
\Omega_{\rm GW}h^2=\Omega_rh^2\frac{d\Omega_{\rm GW}(a_{\rm es})}{d\ln k}\left(\frac{g_0}{g_*}\right)^{1/3},
\end{equation}
where $a_{\rm es}$ is the scale factor at the end of the simulation, $\frac{g_0}{g_*}=\frac{1}{100}$ is the ratio of number of degrees of freedom today to the one at matter-radiation equality and $\Omega_r$ is the radiation density today. Note that the wave number $k$ is related to the frequency $f$ by 
\begin{equation}
f=6\times10^{10}\frac{k}{\sqrt{M_pH_{{\rm ei}}}}{\rm Hz},
\end{equation}
where $H_{{\rm ei}}$ is the Hubble parameter at the end of inflation.

\section{Numerical Simulation and Results}
\label{sec:results} 
To simulate the evolution of coupled scalar fields in an expanding universe, we turn to the public available code PSpectRe~\cite{Easther:2010qz} in which the Fourier-space pseudo-spectral methods are adopted. Here we need to add the evolution of tensor perturbations into PSpectRe without considering the back-reaction of tensor perturbations on the evolution of scalar fields. In simulation, all of the dimensional quantities should be rescaled to be dimensionless as follows 
\begin{eqnarray}
&&\phi_{\rm pr}\equiv Aa^r\phi,~~~\chi_{\rm pr}\equiv Aa^r\chi, \\
&&V_{\rm pr}(\phi_{\rm pr},\chi_{\rm pr})\equiv \frac{A^2}{B^2}a^{-2s+2r}V(\phi,\chi), \\
&&u_{ij,\rm pr}\equiv a^ru_{ij},~~~\vec{x}_{\rm pr}\equiv B\vec{x},~~~dt_{\rm pr}\equiv Ba^sdt.
\end{eqnarray}
Here we consider a simple preheating after $\lambda\phi^4$ chaotic inflation with potential 
\begin{equation}
V(\phi,\chi)=\frac{1}{4}\lambda_{\phi}\phi^4+\frac{1}{2}g^2\phi^2\chi^2,
\end{equation}
where $\lambda_{\phi}=10^{-14}$, $g^2/\lambda_{\phi}=120$ and the initial value of $\phi(k=0)$ is $\phi_0=0.342M_p$. 
By setting the Newton’s gravitational constant $G_N$, the Planck mass $M_p$ and the other rescaling variables $\{A,B,r,s\}$ as
\begin{eqnarray}
\nonumber
&&G_N=1,~~~M_p=1,~~~A=\frac{1}{\phi_0},\\
&&B=\sqrt{\lambda_{\phi}}\phi_0,~~~r=1,~~~s=-1,
\end{eqnarray}
the evolution of scalar fields, background and tensor perturbations in simulation take the following form 
\begin{eqnarray}
\nonumber
\phi''_{\rm pr}&=&a^{-2s-2}\nabla^2_{\rm pr}\phi_{\rm pr}+\left(r(s-r+2)\left(\frac{a'}{a}\right)^2+r\frac{a''}{a}\right)\phi_{\rm pr}\\
&-&\frac{\partial V_{\rm pr}(\phi_{\rm pr},\chi_{\rm pr})}{\partial\phi_{\rm pr}},
\\
\nonumber
\chi''_{\rm pr}&=&a^{-2s-2}\nabla^2_{\rm pr}\chi_{\rm pr}+\left(r(s-r+2)\left(\frac{a'}{a}\right)^2+r\frac{a''}{a}\right)\chi_{\rm pr}\\
&-&\frac{\partial V_{\rm pr}(\phi_{\rm pr},\chi_{\rm pr})}{\partial\chi_{\rm pr}},
\\
\nonumber
\frac{a''}{G_N}&=&(-s-2)\frac{a'^2}{a}+\frac{8\pi}{A^2}a^{-2s-2r-1}\\
&\times&\left(\frac{1}{3}((\nabla_{\rm pr}\phi_{\rm pr})^2+(\nabla_{\rm pr}\chi_{\rm pr})^2)+a^{2s+2}V_{\rm pr}\right),
\\
\nonumber
u''_{ij,\rm pr}&=&\left(r(s-r+2)\left(\frac{a'}{a}\right)^2+r\frac{a''}{a}\right)u_{ij,\rm pr}\\
\nonumber
&+&c_{T}^2a^{-2s-2}\nabla^2_{\rm pr}u_{ij,\rm pr}+\frac{16\pi G_N}{A^2}a^{-2-2s-r}\\
\nonumber
&\times&\bigg(\partial_{i,\rm pr}\phi_{\rm pr}\partial_{j,\rm pr}\phi_{\rm pr}+\partial_{i,\rm pr}\chi_{\rm pr}\partial_{j,\rm pr}\chi_{\rm pr}\\
&-&\left.\frac{1}{3}\delta_{ij}(\partial_{k,\rm pr}\phi_{\rm pr}\partial^k_{\rm pr}\phi_{\rm pr}+\partial_{k,\rm pr}\chi_{\rm pr}\partial^k_{\rm pr}\chi_{\rm pr})\right),
\end{eqnarray}
where primes denote derivative with respect to program time $t_{\rm pr}$. Since the simulation is done in Fourier-space, $\nabla^2_{\rm pr}$ is simply replaced with  $-\vec{k}_{\rm pr}\cdot\vec{k}_{\rm pr}$. The source terms of tensor perturbations should be calculated in position-space firstly using the following simplest discretization scheme
\begin{eqnarray}
\nonumber
\partial_{1,\rm pr}f_{\rm pr}|_{i_1,i_2,i_3}\equiv\frac{1}{2\Delta_{\rm pr}}(f_{\rm pr}|_{i_{1+1},i_2,i_3}-f_{\rm pr}|_{i_{1-1},i_2,i_3}),\\
\nonumber
\partial_{2,\rm pr}f_{\rm pr}|_{i_1,i_2,i_3}\equiv\frac{1}{2\Delta_{\rm pr}}(f_{\rm pr}|_{i_1,i_{2+1},i_3}-f_{\rm pr}|_{i_1,i_{2-1},i_3}),\\
\partial_{3,\rm pr}f_{\rm pr}|_{i_1,i_2,i_3}\equiv\frac{1}{2\Delta_{\rm pr}}(f_{\rm pr}|_{i_1,i_2,i_{3+1}}-f_{\rm pr}|_{i_1,i_2,i_{3-1}}),
\end{eqnarray}
where $\Delta_{\rm pr}\equiv L_{\rm pr}/N=20/256$ is the coordinate distance between neighboring points in position space, and then we transform the whole source terms back to Fourier-space. As for the initial inhomogeneous seeds of the scalar fields, we choose LatticeEasy-style initial conditions~\cite{Felder:2000hq}.

\begin{figure}[]
\begin{center}
\includegraphics[scale=0.25]{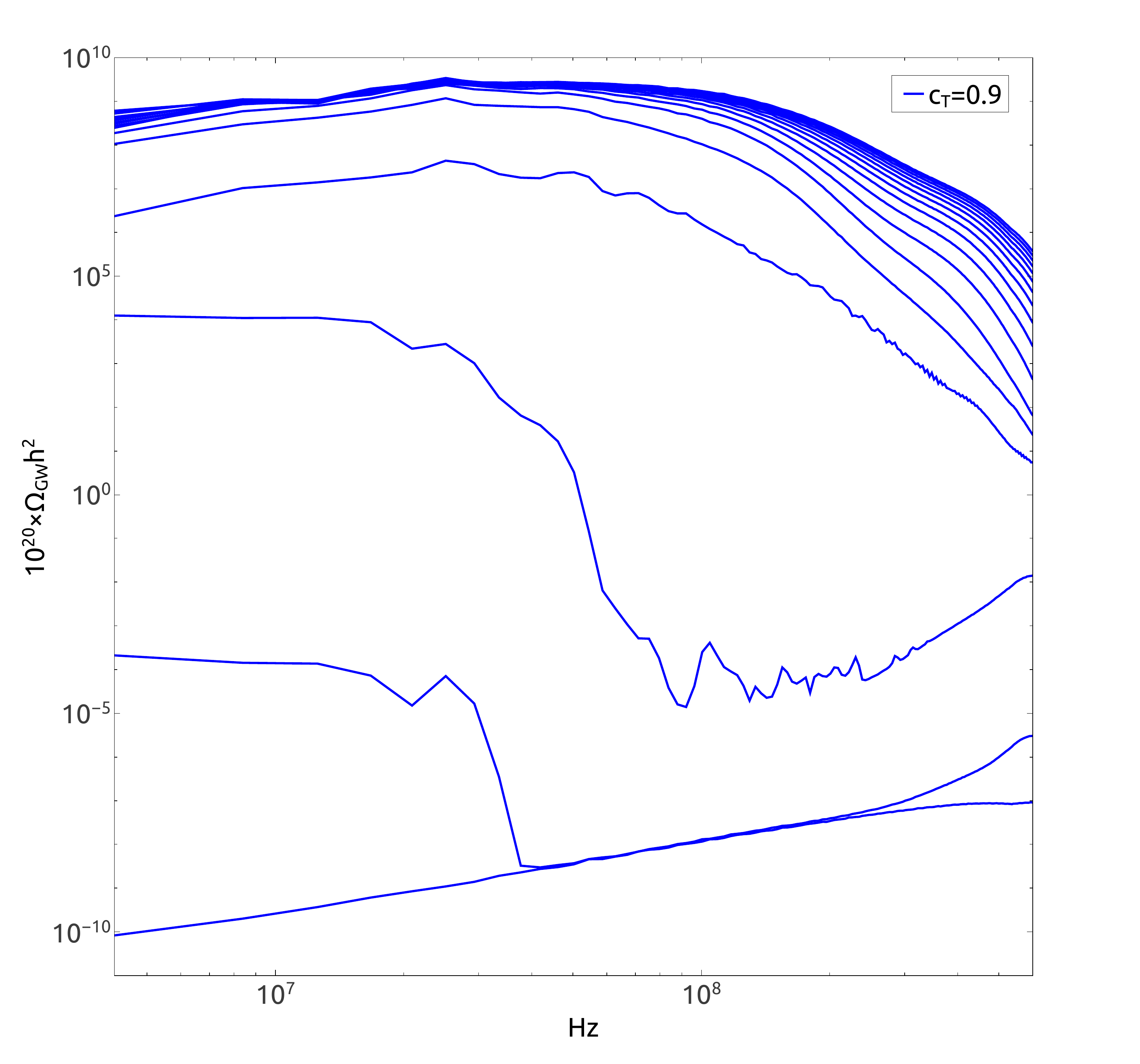}
\includegraphics[scale=0.25]{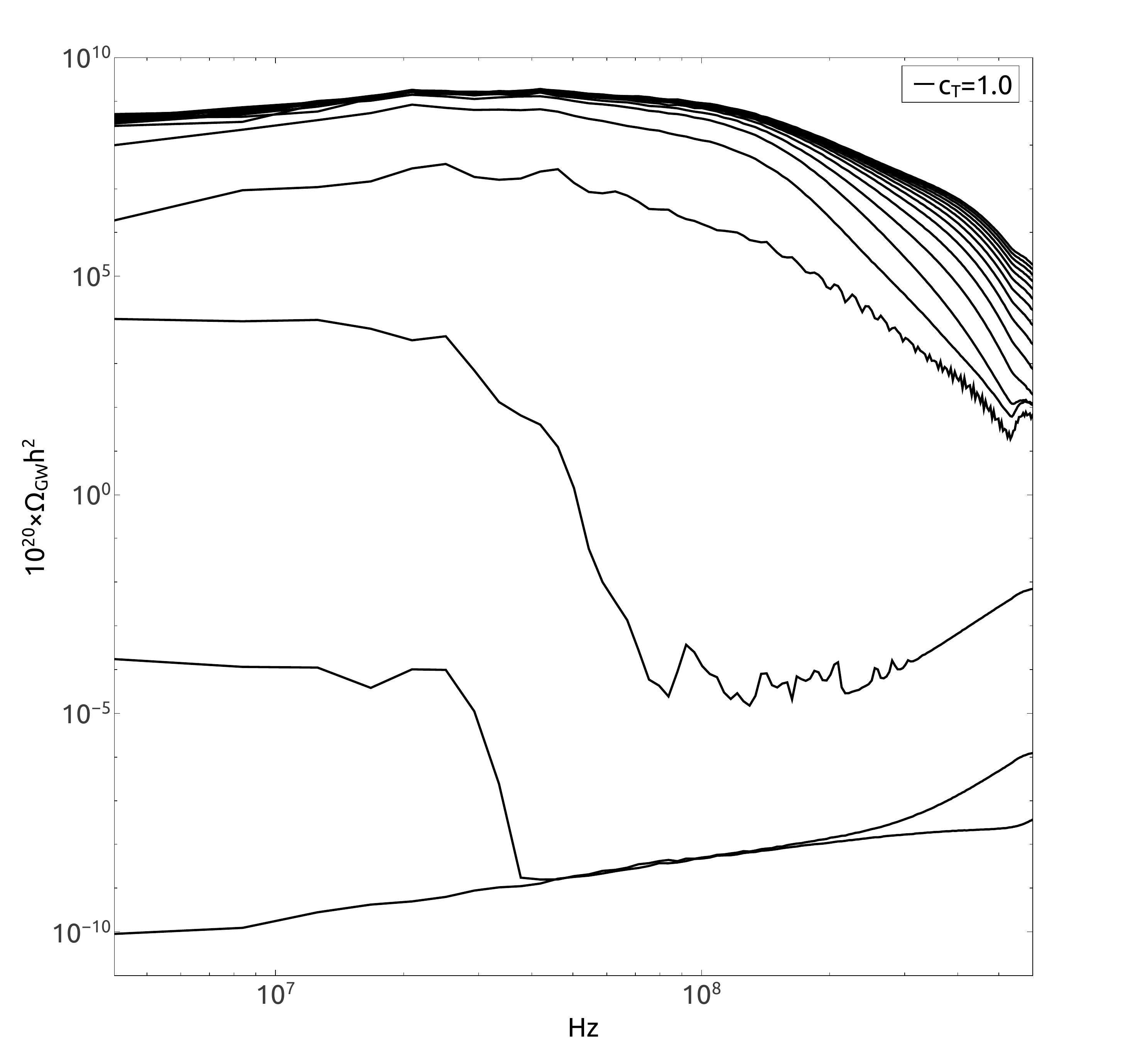}
\includegraphics[scale=0.25]{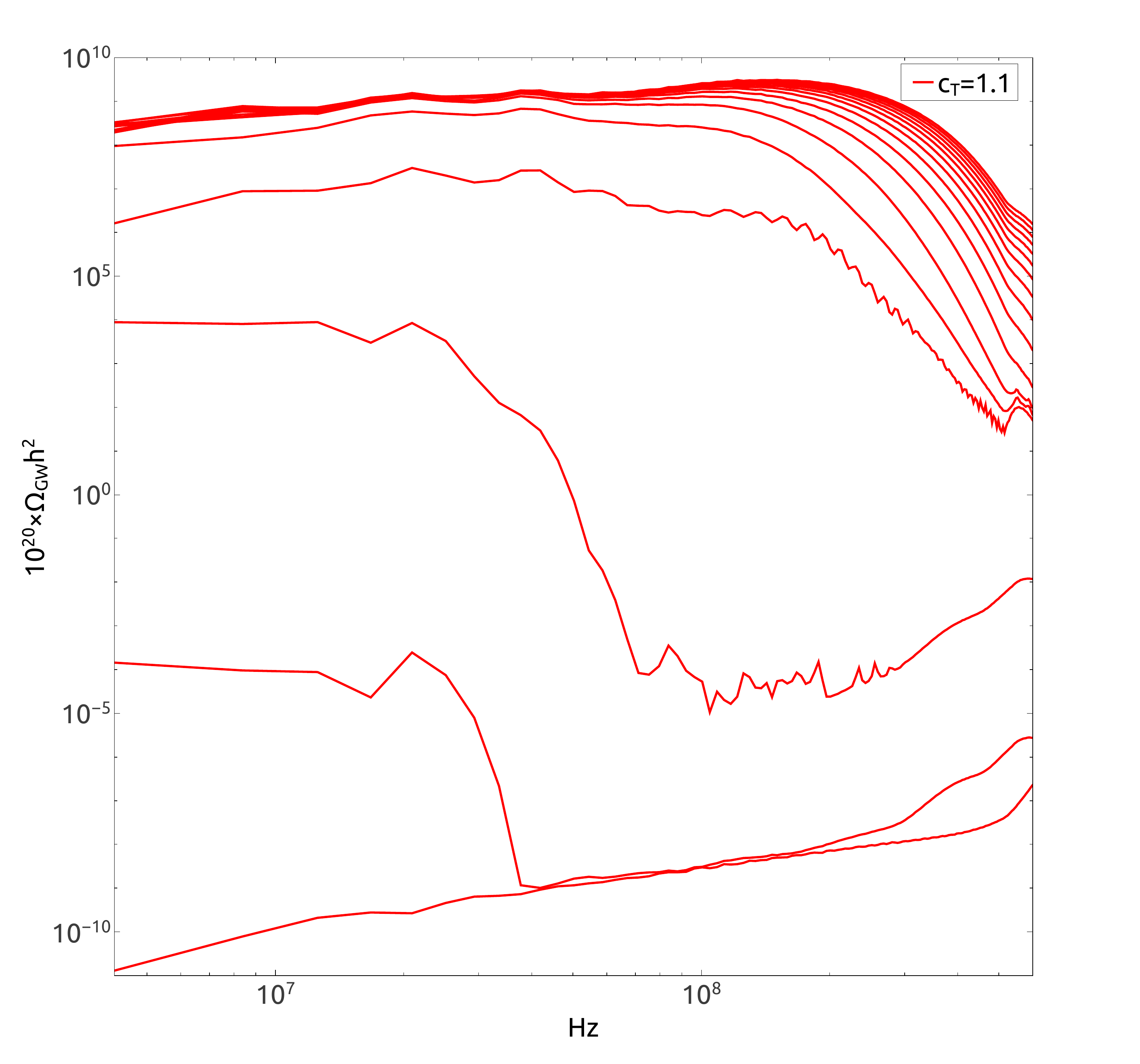}
\end{center}
\caption{The evolution of GW spectra for $c_T=0.9$ (blue), $c_T=1.0$ (black) and $c_T=1.1$ (red) respectively.}
\label{fig:gwev}
\end{figure}

Undoubtedly, $c_T$ would affect the production and evolution of GWs during the whole preheating. Fig.~\ref{fig:gwev} shows the evolution of GW spectra for $c_T=0.9$ (blue), $c_T=1.0$ (black) and $c_T=1.1$ (red) respectively. 
We find that the effect of $c_T$ on GW power spectrum becomes stable at lower frequency first because the inhomogeneity in the field grows faster for smaller $k$ in $\lambda\phi^4$ model.

\begin{figure}[]
\begin{center}
\includegraphics[scale=0.25]{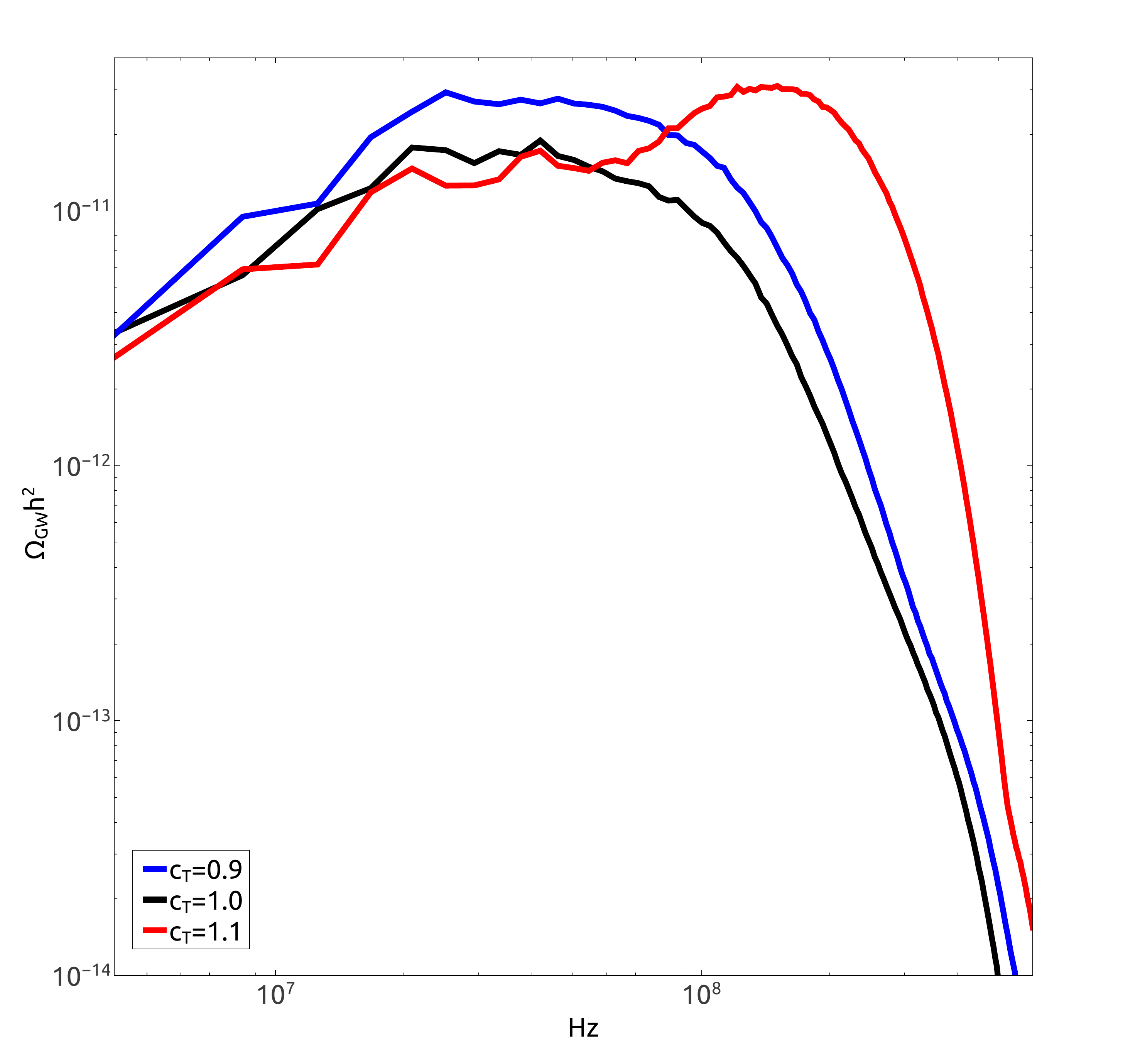}
\end{center}
\caption{Today's GW power spectra for $c_T=0.9$ (blue), $c_T=1.0$ (black) and $c_T=1.1$ (red) respectively.}
\label{fig:gwsp}
\end{figure}

Today's GW power spectra for $c_T=0.9$ (blue), $c_T=1.1$ were shown in Fig.~\ref{fig:gwsp}, respectively.
From Fig.~\ref{fig:gwsp}, it indicates that the GW power spectrum is enhanced for $c_T=0.9$, and the GW power spectrum is distorted for $c_T=1.1$, i.e. the GW power spectrum is suppressed at low frequency but enhanced at high frequency. Actually we find similar behaviors for $c_T<1$ or $c_T>1$ respectively.


\section{Summary and discussion}
\label{sec:sum} 
In this paper, we use Fourier-space pseudo-spectral methods to simulate the evolution of inflaton $\phi$ and scaler field $\chi$ in an expanding universe during preheating after the end of inflation with $\lambda\phi^4$ potential. With the growth of  inhomogeneities, the GWs were generated simultaneously. Once the GW propagation speed $c_T$ is different from the speed of light, today's GW power spectrum is different from that for $c_T=1$. More precisely, the effect of $c_T$ on today's GW power spectrum is a distortion for $c_T=1.1$ but an enhancement for $c_T=0.9$. Therefore, we may use GWs from preheating to explore the GW propagation speed in the future.


The $\lambda\phi^4$ inflation with $c_T\neq1$ is taken as a heuristic model in this paper. Although $c_T$ has been tightly constrained to be very closed to the speed of light by the coalescence of a pair of stellar-mass black holes~\cite{Abbott:2017vtc}, the constraints on $c_T$ from the observations at the very high energy scale are still out of the reach (or at most a weak upper limit was reported~\cite{Raveri:2014eea}). Therefore, it is acceptable that GWs during preheating have  superluminal (or subluminal) speed. 

~

~

~

~

~

~

~

~

~

~

\vspace{5mm}
\noindent {\bf Acknowledgments}

We acknowledge the use of HPC Cluster of Tianhe II in National Supercomputing Center in Guangzhou. We would like to thank Yu Sang for his helpful discussions and advices on this paper. K.~W is supported
by grants from NSFC (grant No. 12005084). Q.~G.~H is supported by grants from NSFC (grant No. 11975019, 11690021, 11991052, 12047503),  Strategic Priority Research Program of Chinese Academy of Sciences (Grant No. XDB23000000, XDA15020701), and Key Research Program of Frontier Sciences, CAS, Grant NO. ZDBS-LY-7009.




\begin{thebibliography}{99}
\frenchspacing
\bibitem{Tristram:2020wbi}
M.~Tristram, A.~J.~Banday, K.~M.~G\'orski, R.~Keskitalo, C.~R.~Lawrence, K.~J.~Andersen, R.~B.~Barreiro, J.~Borrill, H.~K.~Eriksen and R.~Fernandez-Cobos, \textit{et al.}
``Planck constraints on the tensor-to-scalar ratio,''
[arXiv:2010.01139 [astro-ph.CO]].

\bibitem{Traschen:1990sw}
J.~H.~Traschen and R.~H.~Brandenberger,
``Particle Production During Out-of-equilibrium Phase Transitions,''
Phys. Rev. D \textbf{42}, 2491-2504 (1990)

\bibitem{Kofman:1994rk}
L.~Kofman, A.~D.~Linde and A.~A.~Starobinsky,
``Reheating after inflation,''
Phys. Rev. Lett. \textbf{73}, 3195-3198 (1994)

\bibitem{Easther:2006gt}
R.~Easther and E.~A.~Lim,
``Stochastic gravitational wave production after inflation,''
JCAP \textbf{04}, 010 (2006)

\bibitem{Easther:2006vd}
R.~Easther, J.~T.~Giblin, Jr. and E.~A.~Lim,
``Gravitational Wave Production At The End Of Inflation,''
Phys. Rev. Lett. \textbf{99}, 221301 (2007)

\bibitem{GarciaBellido:2007af}
J.~Garcia-Bellido, D.~G.~Figueroa and A.~Sastre,
``A Gravitational Wave Background from Reheating after Hybrid Inflation,''
Phys. Rev. D \textbf{77}, 043517 (2008)

\bibitem{Dufaux:2007pt}
J.~F.~Dufaux, A.~Bergman, G.~N.~Felder, L.~Kofman and J.~P.~Uzan,
``Theory and Numerics of Gravitational Waves from Preheating after Inflation,''
Phys. Rev. D \textbf{76}, 123517 (2007)

\bibitem{Easther:2007vj}
R.~Easther, J.~T.~Giblin and E.~A.~Lim,
``Gravitational Waves From the End of Inflation: Computational Strategies,''
Phys. Rev. D \textbf{77}, 103519 (2008)

\bibitem{Huang:2011gf}
Z.~Huang,
``The Art of Lattice and Gravity Waves from Preheating,''
Phys. Rev. D \textbf{83}, 123509 (2011)

\bibitem{Giblin:2011yh}
J.~T.~Giblin, Jr., L.~R.~Price, X.~Siemens and B.~Vlcek,
``Gravitational Waves from Global Second Order Phase Transitions,''
JCAP \textbf{11}, 006 (2012)

\bibitem{Zhou:2013tsa}
S.~Y.~Zhou, E.~J.~Copeland, R.~Easther, H.~Finkel, Z.~G.~Mou and P.~M.~Saffin,
``Gravitational Waves from Oscillon Preheating,''
JHEP \textbf{10}, 026 (2013)

\bibitem{Ashoorioon:2013oha}
A.~Ashoorioon, B.~Fung, R.~B.~Mann, M.~Oltean and M.~M.~Sheikh-Jabbari,
``Gravitational Waves from Preheating in M-flation,''
JCAP \textbf{03}, 020 (2014)

\bibitem{Liu:2017hua}
J.~Liu, Z.~K.~Guo, R.~G.~Cai and G.~Shiu,
``Gravitational Waves from Oscillons with Cuspy Potentials,''
Phys. Rev. Lett. \textbf{120}, no.3, 031301 (2018)

\bibitem{Amin:2018xfe}
M.~A.~Amin, J.~Braden, E.~J.~Copeland, J.~T.~Giblin, C.~Solorio, Z.~J.~Weiner and S.~Y.~Zhou,
``Gravitational waves from asymmetric oscillon dynamics?,''
Phys. Rev. D \textbf{98}, 024040 (2018)

\bibitem{Adshead:2018doq}
P.~Adshead, J.~T.~Giblin and Z.~J.~Weiner,
``Gravitational waves from gauge preheating,''
Phys. Rev. D \textbf{98}, no.4, 043525 (2018)

\bibitem{Liu:2018rrt}
J.~Liu, Z.~K.~Guo, R.~G.~Cai and G.~Shiu,
``Gravitational wave production after inflation with cuspy potentials,''
Phys. Rev. D \textbf{99}, no.10, 103506 (2019)

\bibitem{Sang:2019ndv}
Y.~Sang and Q.~G.~Huang,
``Stochastic Gravitational-Wave Background from Axion-Monodromy Oscillons in String Theory During Preheating,''
Phys. Rev. D \textbf{100}, no.6, 063516 (2019)

\bibitem{Jin:2020tmm}
G.~Jin, C.~Fu, P.~Wu and H.~Yu,
``Production of gravitational waves during preheating in the Starobinsky inflationary model,''
Eur. Phys. J. C \textbf{80}, no.6, 491 (2020)

\bibitem{Hiramatsu:2020obh}
T.~Hiramatsu, E.~I.~Sfakianakis and M.~Yamaguchi,
``Gravitational wave spectra from oscillon formation after inflation,''
[arXiv:2011.12201 [hep-ph]].

\bibitem{Sang:2020kpd}
Y.~Sang and Q.~G.~Huang,
``Oscillons during Dirac-Born-Infeld Preheating,''
[arXiv:2012.14697 [hep-th]].





\bibitem{Saltas:2014dha}
I.~D.~Saltas, I.~Sawicki, L.~Amendola and M.~Kunz,
``Anisotropic Stress as a Signature of Nonstandard Propagation of Gravitational Waves,''
Phys. Rev. Lett. \textbf{113}, no.19, 191101 (2014)

\bibitem{Misner:1973}
C.~W.~Misner, K.~S.~Thorne and J.~A.~Wheeler,
``Gravitation,''
San Francisco, 1973, 1279p

\bibitem{Easther:2010qz}
R.~Easther, H.~Finkel and N.~Roth,
``PSpectRe: A Pseudo-Spectral Code for (P)reheating,''
JCAP \textbf{10}, 025 (2010)

\bibitem{Felder:2000hq}
G.~N.~Felder and I.~Tkachev,
``LATTICEEASY: A Program for lattice simulations of scalar fields in an expanding universe,''
Comput. Phys. Commun. \textbf{178}, 929-932 (2008)

\bibitem{Abbott:2017vtc}
B.~P.~Abbott \textit{et al.} [LIGO Scientific and VIRGO],
``GW170104: Observation of a 50-Solar-Mass Binary Black Hole Coalescence at Redshift 0.2,''
Phys. Rev. Lett. \textbf{118}, no.22, 221101 (2017)
[erratum: Phys. Rev. Lett. \textbf{121}, no.12, 129901 (2018)]

\bibitem{Raveri:2014eea}
M.~Raveri, C.~Baccigalupi, A.~Silvestri and S.~Y.~Zhou,
``Measuring the speed of cosmological gravitational waves,''
Phys. Rev. D \textbf{91}, no.6, 061501 (2015)

\end{thebibliography}
\end{document}